# First Principles Study of the Impact of Grain Boundary Formation in the Cathode Material LiFePO₄

Jan Kuriplach [1,*], Aki Pulkkinen [2], and Bernardo Barbiellini [2,3]

[1] *Department of Low Temperature Physics, Faculty of Mathematics and Physics, Charles University, V Holešovičkách 2, CZ-180 00 Prague, Czech Republic*
[2] *Department of Physics, School of Engineering Science, LUT University, FI-53851, Lappeenranta, Finland*
[3] *Physics Department, Northeastern University, Boston, MA 02115, USA*
[*] Correspondence: Jan.Kuriplach@mff.cuni.cz

**Abstract:** Motivated by the need to understand the role of internal interfaces in Li migration occurring in Li-ion batteries, a first principles study of a coincident site lattice grain boundary in LiFePO₄ cathode material and in its delithiated counterpart FPO₄ is performed. The structure of the investigated grain boundary is obtained and the corresponding interface energy is calculated. Other properties, such as ionic charges and magnetic moments, excess free volume and the lifetime of positrons trapped at the interfaces, are determined and discussed. The results show that while the grain boundary in LiFePO₄ has desired structural and bonding characteristics, the analogous boundary in FePO₄ needs to be yet optimized to allow for an efficient Li diffusion study.

## 1. Introduction

A key challenge for developing efficient lithium-ion batteries (LIBs) is to preserve homogeneous Li flows. Li inhomogeneity can appear at internal interfaces within the LIB. Grain boundaries that result from heat treatments aimed to improve the phase crystallinity of LIB cathodes can therefore hinder the homogeneity of the Li distribution. Among LIB cathodes, LiFePO₄ or triphylite (aka lithium iron phosphate, LFP) is an environmentally friendly and cheap material [1]. It has an orthorhombic structure [1,2], where the Li atoms occupying octahedral lattice sites diffuse in and out while the LIB works. A major challenge for LIB durability is ageing produced by several factors that have a complex hierarchical structure on various length and time scales. Due to these issues, in spite of several decades of research, the understanding of battery ageing remains challenging. Recent investigations [3-7] aimed at studying LFP have combined x-ray spectroscopy and theoretical modeling to monitor the evolution of the redox orbitals in nanoparticles and single-crystal LFP cathodes under different lithiation levels. These studies have provided advanced characterizations techniques for cathodes such as general methods for understanding the relation between lattice distortions and potential shifts [6]. Recent review [8] summarizes the status and perspectives of LIB cathodes based on the olivine structure, including LFP. Since LIB cathodes are made of powders, grain surfaces and interfaces come to play. Grain boundaries (GBs) – *i.e.* interfaces between the grains of the same material -- are the simplest interfaces that can be studied using reliable density functional theory (DFT) calculations. Recently Lachal *et al.* [9] have surmised that strong chemical delithiation kinetic degradation in LFP can be explained by grain boundaries that drastically hinder the lithium mobility via the obstruction of the phase propagation, which may cause stress at the GBs and crack formation. In this paper, we present a theoretical study based on DFT calculations aimed to be a first step in clarifying this hypothesis. In particular, we examine one type of GB both for lithiated and delithiated systems, explore various GB properties, and discuss differences between the GB in LFP and its delithiated equivalent. This constitutes the basis for a future study of Li diffusion at the GBs in LFP. Li diffusion at GBs represents another aspect of the GB effect on the LIB operation [10] since the diffusion along and across the GB may substantially differ.

**2. Grain Boundary Construction and Computational Methods**

The triphylite exhibits an orthorhombic structure with the space group *Pnma* (No. 62, standard setting) [2]. In this case, the lattice parameters for triphylite are in the order $a > b > c$, and the structure mirror planes' normal is along the axis ***b***. Wyckoff positions of individual atoms are: Li – 4a, Fe – 4c, P – 4c, O(1) – 4c, O(2) – 4c, and O(3) – 8d; *i.e.* 28 atoms per unit cell in total (Z = 4). There are three different types of oxygen positions differing in their tetrahedral cationic coordination. The coordination number of both Fe and Li ions is 6, *i.e.* there is a distorted octahedron made of 6 oxygen anions. In the case of delithiated structure, $FePO_4$ (heterosite, iron phosphate, FP), Li atoms are missing, but the structure type and space group remain the same (the unit cell volume diminishes compared to LFP). The Fe ion coordination type does not change with respect to LFP, but oxygen coordination is changed drastically being 2- or 3-fold. The Fe ion valence is considered to be nominally 2+ in LFP (LIB discharged/lithiated) and 3+ in FP (LIB charged/delithiated) systems.

There can be many types of GBs for a given material. We will consider just planar GBs with suitable periodic boundary conditions that can be handled with *ab initio* electronic structure codes (classical molecular dynamics could in principle be performed too if appropriate interatomic potentials are available). A geometrical method based on the 'coincidence site lattice' (CSL) principle [11] allows us to construct various GB boxes suitable for *ab initio* calculations. Technically, the method consists in cutting the crystal along a crystallographic plane and rotating one part of the crystal along a chosen axis in the way that some lattice sites of the rotated and unrotated crystals coincide, creating thus a lattice of coincident sites, *i.e.* a CSL. When the rotation axis is laying in (perpendicular to) the crystallographic plane considered, the obtained GB is called tilt (twist). Applying such a CSL construction method yields a first approximation to the GB atomic structure and the relaxation of atomic sites at the GB and its surrounding is necessary. In the following, we will consider a tilt GB in LFP and FP. The CSL concept works well for cubic systems where the coincidence (in CSL sense) can be made exact. However, this is not the case of lower symmetry structures like orthorhombic (L)FP (see Ref. [12]). For such cases, the coincidence can be made approximate only and the corresponding GB is then called a 'near-CSL' GB.

The relaxation of GB configurations was performed using the Vienna ab-initio simulation package (VASP) [13,14] with the projector-augmented wave (PAW) method [15] and the Perdew-Burke-Ernzerhof (PBE) exchange-correlation functional [16] within the spin-polarized generalized gradient approximation (GGA) [17]. PAW potentials [14] with valence electron configurations $2s^1$ for lithium, $3p^6\ 3d^7\ 4s^1$ for iron, $3s^2\ 3p^3$ for phosphorus and $2s^2\ 2p^4$ for oxygen were used. The energy cutoff was set to 400 eV in calculations involving cell dimension relaxations, whereas 300 eV was employed in cases where the cell size remained fixed. The cell shape is given by the GB geometrical construction outlined above. A 2×2×2 Monkhorst-Pack ***k***-point grid and Gaussian smearing with the width 0.1 eV was used to sample the Brillouin zone. The criterion for the energy convergence was 0.1 meV and in geometry optimizations the forces were converged within 10 meV/Å. The starting values of lattice parameters and atomic coordinates were taken from Refs. [2] and [18] for LFP and FP systems, respectively. Since LFP and FP are antiferromagnetic (AF) at low temperatures and paramagnetic at LIB operating temperatures, the spin-polarization of (L)FP was explicitly taken into account in relaxations by considering an AF order of iron ions. More specifically, spin-polarization has to be taken into account since the non-magnetic LFP unit cell relaxations lead to an equilibrium cell volume which is about 9% smaller than the experimental one whereas magnetic calculations (ferromagnetic and antiferromagnetic) lead to a volume which is just 2% larger than experiment. The antiferromagnetic arrangement gives the lowest energy for the bulk materials. A similar situation occurs for the FP case when the non-magnetic state exhibits almost 10 % larger volume, but the magnetic ones results in 6 – 8 % larger volumes, which is not optimal, but we need to handle both LFP and FP systems on equal footing. The optimization of supercells with respect to atomic positions is quite time consuming because of charge sloshing effects complicated further by the AF order. The total energies of relaxed supercells served to calculate the GB energies. For this purpose, the geometry of a bulk cell with the same number of atoms and shape as the GB cell was also optimized (both for LFP and FP cases), and the corresponding total energy was taken as a reference. In order to assess ionic charges and magnetic moments, the so-called Bader charge analysis [19] was performed using the implementation [20] for VASP.

Positron annihilation spectroscopy [21] can, in principle, be used to study materials defects possessing some open volume, like grain boundaries. In order to check how the obtained GB configurations interact with positrons – in particular, whether they get trapped in the GB region – calculations of the positron lifetime and energy were performed. The computational scheme follows the procedure developed in Ref. [22] with the self-consistent electron density and Coulomb potential obtained by means of the WIEN2k code [23] where GB configurations obtained by VASP were taken. The electron-positron enhancement factor and positron correlation potential were considered within a parameter-free GGA [24] and a local density approximation (LDA) [25] for electron-positron correlations. The real-space code [22] employed to calculate positron characteristics was extended to allow the treatment of non-orthogonal cells with GBs (the details will be published elsewhere). Theoretical positron GB studies are based on those performed already earlier in metals [26] and oxides [26,27].

## 3. Results and Discussion

### 3.1. Lithiated system

The GB construction starts by selecting the tilt axis. We take [010] (perpendicular to the structure mirrors with the coordinates $y = ¼$ and $¾$), which is in fact parallel with the LFP crystallographic axis *b*. Such an axis allows us to construct GBs with Li atoms at the GB plane in order to be able to study Li diffusion at this GB and its neighborhood in a follow-up study. By trying various tilt/rotation angles along the chosen axis selecting the Li atom at the (0,0,0) position as origin, a nearly perfect coincidence occurs at ~48.3° for a near-CSL GB with a {101} plane. The CSL is determined by translation vectors *a-c*, *b*, and 3*c*, which implicates that the CSL unit cell volume is three times more than that of the original lattice determined by vectors *a*, *b*, and *c*. Thus, the complete designation is (near-CSL) symmetrical Σ 3 (101)/[010] GB, with Σ specifying the ratio of volumes. The total volume of the GB box is then 6 times the LFP unit cell volume (the lower part is not rotated and the upper part is rotated by 48.3° degrees along the [010] axis). In this way, each box contains two GB boundaries (one is in the middle and the other one at the upper and lower faces). It is useful to note that Li ions lie on the (101) GB plane and since this plane is not a mirror, the GB lacks mirror symmetry. The result of the construction procedure called further configuration 1 (or C1) is shown in Figure 1a. The GB box (unit cell/supercell) is not anymore orthogonal and contains 168 atoms. In fact, Figure 1a presents the structure whose last dimension (along the original lattice vector 6*c*) is already relaxed; atomic positions in the whole box are also allowed to relax to minimize the total energy. A closer inspection unveils that bonds of atoms near the interface are somewhat distorted and dangling bonds are also present. Especially, four pairs of Fe atoms become mutually very close, which results in their repulsion because of differently positioned O atoms that do not screen well such repulsion compared to the bulk regions. The octahedral coordination of Fe (and Li) atoms is lowered in the GB regions (though one could also see it as a disruption of oxygen cationic coordination).

Since the GB construction procedure described above is based on a geometrical concept, it might not necessarily lead to the best chemical bonding of atoms at the interface (after 'cutting' and 'welding' two pieces of crystal together). In order to bond better the atoms at the GB regions, we shift the upper grain along the [010] direction (*b* axis) by *b*/2, which is what a detailed examination of the interfaces suggests. In this way, the coherence/'matching' of both interfaces could be improved. No atoms are added or removed in the present GB configuration constructed to be charge neutral (the same holds for C1). The result of this transformation (including the relaxation of atoms and the last cell dimension) is shown in Figure 1b and it is called configuration 2 (or C2). From the viewpoint of the magnetic structure, we keep the magnetic moments opposite for Fe atoms with $y = ¼$ and $¾$, having effectively an AF structure. In the relaxed structure C2, the coordination number of Fe atoms close to the GB reaches 4 – 6, whereas the C1 structure yields 4 – 5 only. This means that some Fe atoms close to the interfaces in C2 exhibit the same (octahedral) coordination as in the bulk regions. The oxygen coordination is also modified at the GB region and increased slightly upon the shift performed to obtain configuration C2. The coordination of P ions is not affected at the GB regions. Regardless of configuration, Li atoms are 5-fold coordinated at all GB interfaces. These findings indicate that the presence of cations at the studied GBs is important in order to maintain cohesion at these interfaces.

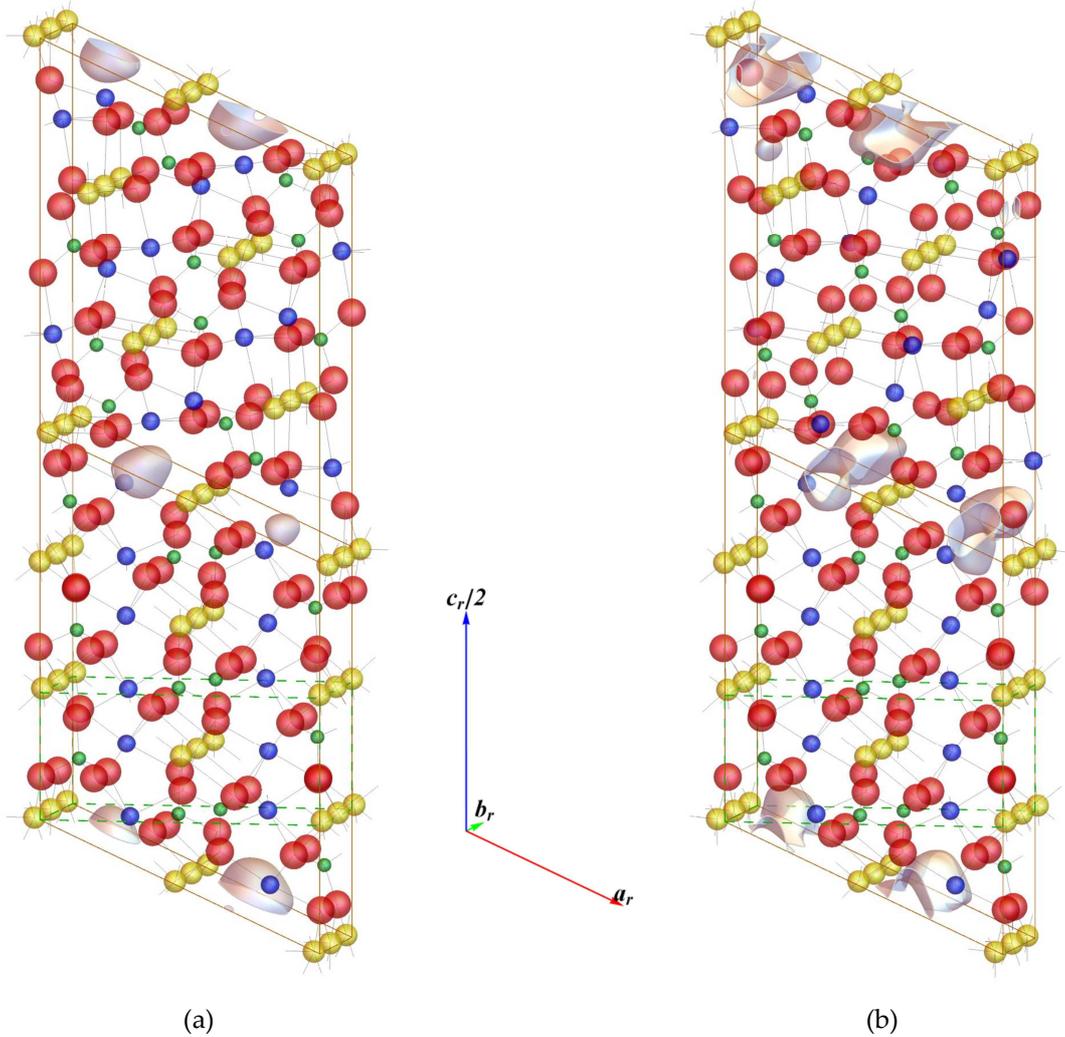

(a)             (b)

**Figure 1.** Σ 3 (101)/[010] GB for LFP: (**a**) configuration C1 ; (**b**) configuration C2. Color scheme is as follows: Li-yellow, Fe-blue, P-green, O-red; the LFP unit cell is also shown as a dashed green cuboid. The GBs in the middle of the boxes are indicated by tilted rectangles. The supercell translation vectors are also outlined. The grayish objects identify the regions with the highest positron density (see the text for explanations).

The final C1 and C2 structures (shown in Figures 1a and 1b) were both relaxed with respect to atomic coordinates as well as the supercell last dimension. This procedure is now explained in more detail as follows. The 'bulk' supercell (not shown here), in which the upper part/grain is not rotated, is relaxed with respect to its all cell dimensions (including atomic positions). The corresponding lattice vectors are $a_r$, $b_r$, and $c_r$ and it holds approximately that $a_r \approx a - c$, $b_r \approx b$, and $c_r \approx 6c$. In this way, we obtain the supercell containing no GB/interface and this supercell is used as a reference for GB energy calculations (corresponding total energy is denoted as $E_{tot}(bulk)$). First two box dimensions along cell vectors $a_r$ and $b_r$ are taken to be the corresponding dimensions of supercells for configurations C1 and C2. The last cell dimension, called further $z_c$, was taken initially to be $c_r$ and its size is then optimized to obtain the lowest energy, $E_{tot}(GB)$. The final/equilibrium value of $z_c$ is larger than $c_r$ since the cells expand along the last cell vector (the adaptation of the structure at the grain interfaces takes normally larger volume compared to bulk). Whereas dimensions parallel to the interface are not changed as the same type of periodicity is kept as in the bulk. The curves showing how the cells' energy is changed with the dimension $z_c$ are presented in Figure 2. In fact, the grain boundary energy, γ, is plotted here against $z_c$ for both GB configurations, and the minimum corresponds to equilibrium, *i.e.* the optimal dimension. The GB energy (or interface energy) is given by relation

$$\gamma = [\, E_{tot}(GB) - E_{tot}(bulk)\,]\, /\, 2A, \qquad (1)$$

where $A$ is the area of the GB in the supercell/box, considering that there are two GBs in each supercell. Using the dimensions $a_r$, $b_r$, and $c_r$ of the relaxed bulk supercell, $A$ can be calculated as $a_r b_r \approx b(a^2+c^2)^{1/2}$ when we relate it to the original primitive lattice dimensions. The minimum along this relaxation path corresponds to the equilibrium volume. Our results on the GB interface energy are given in Table 1. One can see that the GB energies are quite low compared to a typical GB energy around 1 J/m², which indicates that the studied interfaces should be relatively easy to form. It also holds that γ(C1) > γ(C2) showing that the configuration C2 is more energetically favorable than C1 even if the difference is smaller than 10 %. This is an expected result since C2 exhibits better coherence than C1, as discussed above. Table 1 contains γ values both in J/m² and meV/Å² units since the latter is more suitable for considerations at the atomic level (1 J/m² corresponds to 62.4 meV/Å²). In principle, the GB energy can be recalculated to the excess energy per one atom/ion at the interface (here ~0.5 eV/Li), which could be related, for example, to the point defect formation energies that are of the same order.

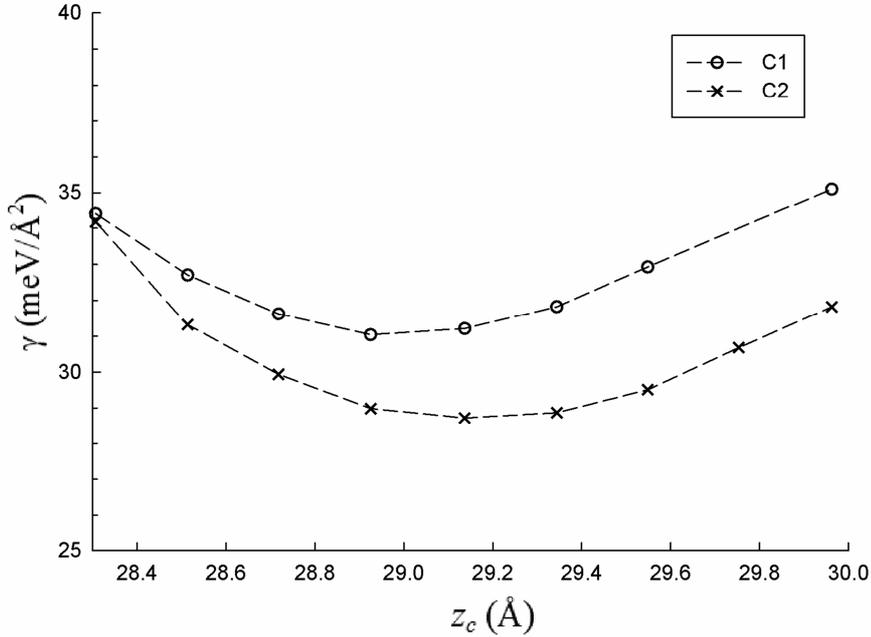

**Figure 2.** Relaxation of the LFP GB boxes along the last cell vector for configurations C1 and C2: grain boundary energy (γ) versus box last dimension ($z_c$).

Another observation is that C2 expands more than C1 as it is reflected by the GB excess free volumes $\delta V(C1) = 0.26$ Å³/Å² and $\delta V(C2) = 0.36$ Å³/Å². This quantity can be determined as an extra volume due to the introduction of the GB into the bulk related to the GB area (units are usually expressed in the form volume per area), and can be easily calculated using Equation (1) where energies are replaced by cell volumes. The property $\delta V$ should be viewed as an increase in the interstitial space at GB regions. GB excess free volumes can be also connected to positron characteristics, especially the lifetime. The bulk positron lifetime for the LFP structure was calculated in Refs. [28,29] and amounts to 157.5 ps (LDA) and 169.7 ps (GGA) considering room temperature lattice parameters. Here we are getting [30] slightly longer values 159 ps and 171 ps for respective treatments of electron-positron correlations considering the relaxed lattice parameters for which the volume per formula unit is larger (and consequently is the positron lifetime). We give both LDA and GGA values, even if the latter one is more realistic for comparison with experiment, to evaluate the effect of gradient corrections which is apparently quite large because of open type crystal structure and ionic character of interatomic bonds. As for GBs, Figures 1a and 1b show clearly – positron density isosurfaces at about 55 % of the maximum value are displayed – that positron presence is enhanced at GB regions, especially in interstitial channels between Li chains along the [010] direction. This is also reflected by the positron lifetime, $\tau_{GB}$, which is 10–20 ps longer than its bulk counterpart. In particular, we obtain $\tau_{GB}$ = 177 ps (LDA) and 190 ps (GGA) for C1, whereas $\tau_{GB}$ = 169 ps (LDA) and 182 ps (GGA) for C2. Small differences in positron density distributions between the middle and bottom/top GB are likely due to not exact site coincidence at GBs mentioned above. Surprisingly, $\delta V$(C2) > $\delta V$(C1) but $\tau_{GB}$(C2) < $\tau_{GB}$(C1). In principle, a larger (excess) free volume should result in a larger positron lifetime, but in this case a complex morphology of the GB interstitial space also plays a role, which likely reverses the expected trend (compare Figures 1a and 1b). In any case, we can state that positrons trap at the studied GB in the LFP material. Interestingly, recent experiment [31] provides positron lifetime data of a LiFePO$_4$ powder sample with the average grain size about 100 nm. A lifetime component of 188 ps was detected and assigned to the annihilation of free positrons. Considering the above calculated values of the bulk positron lifetime, we rather suggest that the component 188 ps more probably corresponds to the annihilation at grain boundaries because of similarity with lifetimes for GBs calculated here (GGA case). Clearly, more types of GBs need to be examined to see how the lifetime depends on the GB geometry. Positrons stay in the interstitial space, but they are affected by ions forming an interstitial space 'boundary'. Positrons are attracted to negative ions and repulsed from the positive ones.

Table 1. GB interface energy for LFP and FP configurations.

| GB configuration | $\gamma$ (meV/Å$^2$) | $\gamma$ (J/m$^2$) |
|---|---|---|
| LFP C1 | 31 | 0.50 |
| LFP C2 | 29 | 0.46 |
| FP  C1 | 69 | 1.11 |
| FP  C2 | 58 | 0.92 |

The atom/ion charge analysis after Bader [19] allows to assign charges to ions in a physically plausible way using the concept of 'zero flux surface' of the electron density around ions. Bader charges then correspond to the electron density integrated inside the (Bader) volumes bounded by zero flux surfaces. Using the implementation [20] for VASP, we found that in the bulk cell average charges, $q$, related to neutral atoms for all atomic/crystallographic species are: $q$(Li) = +0.88$e$, $q$(Fe) = +1.40$e$, $q$(P) = +3.63$e$, $q$(O(1)) = −1.50$e$, $q$(O(2)) = −1.50$e$, and $q$(O(3)) = −1.46$e$ (with $e$ being the elementary charge). Summing them for the formula unit, *i.e.* $q$(Li) + $q$(Fe) + $q$(P) + $q$(O(1)) + $q$(O(2)) + 2$q$(O(3)), gives −0.01$e$, which is effectively zero considering the rounding errors. The charge neutrality is preserved, as expected, and the charges are not too far from nominal valence charges assumed for LFP (note that the phosphate group, nominally (PO$_4$)$^{3-}$, has charge −2.29$e$). Concerning the GB configuration C1, the average ionic charges almost do not change (at most 0.02$e$ in both directions), but the charge of Fe ions close to GB interfaces is lowered by 0.14$e$ in average (there are 8 such Fe ions). For other atomic species, the changes are apparently smaller though they must compensate lowered Fe charges in the whole cell. The detailed examination of atomic clusters making 'formula units' (*i.e.* LiFePO(1)O(2)O(3)$_2$) reveals that both GB interfaces in the cell are slightly negatively charged. In particular, atomic clusters close to the interface have charge between −0.12$e$ and −0.16$e$, whereas clusters in the bulk regions have about half of these charges with the opposite sign. In the case of configuration C2, the charges of Fe ions close to the interface are almost unchanged compared to those in bulk and the effect of GB charging is almost negligible (about –0.02$e$ in average per cluster). This shows that a better matching in bonds, appearing for the configuration C2, can make charges almost the same as in the perfect crystal. Another aspect is that positrons are attracted to negatively charged regions (GBs). This effect may affect the positron lifetime already discussed though a detailed analysis is needed. Similarly to charge, one can integrate spin up and spin down charge densities in the same Bader volumes to get magnetic moments (μ): μ(Li) = 0.00$\mu_B$, μ(Fe) = 3.59$\mu_B$, μ(P) = 0.00$\mu_B$, μ(O(1)) = 0.04$\mu_B$, μ(O(2)) = 0.05$\mu_B$, and μ(O(3)) = 0.03$\mu_B$ ($\mu_B$ is the Bohr magneton; only magnitudes are given). The net magnetic moment of the supercell is zero, as required for an AF order. The magnetic moment of Fe is close to the anticipated value 4$\mu_B$ (spin only contribution). Magnetic moments of oxygen ions are nearly negligible. Concerning the GB configurations, both C1 and C2 have average Fe and O magnetic moments nearly the same as in bulk within 0.02$\mu_B$. In the case of C1, Fe atoms close to the interface have their magnetic moments about 0.12$\mu_B$ smaller than those farther from the interfaces (magnitude is considered). This is related to the effect of diminished charge at such Fe ions. When C2 is considered, Fe atoms with 6-fold coordination at the GB interfaces have their magnetic moment with the same value as in bulk regions, but those with a lower coordination number have their moment lower by about 0.02$\mu_B$ compared to the average value. This indicates that a better coherency of C2 is also detectable via the magnetic moments. The net magnetic moment of the whole supercell is negligible (~0.01$\mu_B$) for both configurations confirming an AF magnetic state.

### 3.2. Delithiated system

In the case of delithiated system, *i.e.* FPO$_4$ or FP, we proceed in a similar way to that employed for LFP. Even if the lattice parameters *a*, *b*, and *c* do not change much when going from LFP to FP by removing Li atoms, the *c*/*a* ratio important for CSL construction varies more significantly (increases by ~7 %). This change affects the tilt angle becoming now ~51.4° and increases the Σ parameter to 4 since Σ 3 yields poor coincidence (certainly tilt angle and Σ are correlated). The result is a near-CSL symmetrical tilt Σ 4 (101)/[010] GB for FP [32]. For this reason, there are more atoms in the boxes/supercells than before, even if Li is not present. The supercells now contain 192 atoms (32× Fe, 32× P, 128× O). Nevertheless, a comparison of constructed GBs in LFP and FP still makes sense since the GB planes (and tilt axes) are the same and thereby the geometrical relationship of the lower and upper blocks (expressed by the tilt angle) is very similar in both cases. However, we have no atoms/cations lying on the GB planes – in contrast to the LFP case, which may pose some problems related to the GB cohesion. In any case, this GB model for FP may be ideal to study Li GB diffusion at a later stage. As in the case of LFP, we construct a GB configuration shifted along [010] by *b*/2 (configuration 2) in order to improve the GB coherence compared to the configuration 1 obtained just by rotation along the tilt axis. The perfect bulk box was constructed and relaxed as well and its $a_r$, $b_r$, and $c_r$ translation vectors were used to make initial configurations of GBs like for LFP. The magnetic order was again considered to be antiferromagnetic and the directions of magnetic moments of Fe atoms were arranged in the same way as in LFP. While checking the atomic neighborhood of interfaces, one can notice that rebonding of atoms at the bottom/top interface might be impeded by the fact that the matching plane of the rotated part is shifted by $a_r$/2, which does not happen for the middle interface. This is an effect of having a Σ 4 GB for FP and it does not happen for the Σ 3 GB studied in LFP.

Figure 3 displays *en face* the relaxed GB configuration 1 (C1) and configuration 2 (C2) of FP without performing box dimension optimization (it has length $c_r$ as in the bulk box). The reason is that we need to examine first the GB structure before proceeding further given the difficulty with the rebonding process mentioned above. Configuration 1 (Figure 3a) clearly has no bonds across the interfaces showing also large interstitial space at these regions. In contrast, configuration 2 exhibits Fe-O bonding across the middle GB whereas bonding across the bottom/top GB appears weak. This might indicate an insufficient GB cohesion in FP, at least for the GB type studied. Preliminary calculations show that increasing the last box dimension leads to upper and lower grain separation at both interfaces for C1. In the case of C2, only the bottom/top interface splits. Concerning the coordination of Fe atoms close to the interfaces, C1 shows 4- or 5-fold coordination at both interfaces, whereas in C2 Fe is 4- or 6-fold coordinated in the middle and 4 – 6 at the bottom/top, which is similar to that observed in LFP. GB energies are calculated formally according to Equation (1). The results are reported in Table 1 and provide the first idea about interface energies for the delithiated system. These values correspond to boxes with the last dimension unrelaxed. We can see that C1 has a larger interface energy compared to C2, which is again expected trend. Both $\gamma$'s – to be considered as a first guess only – are about twice larger than those for LFP. Obviously, the FP GB energies may still change when the studied GB is further elaborated to be better suitable for the Li diffusion study envisaged. Some ideas in this direction are given in the next section.

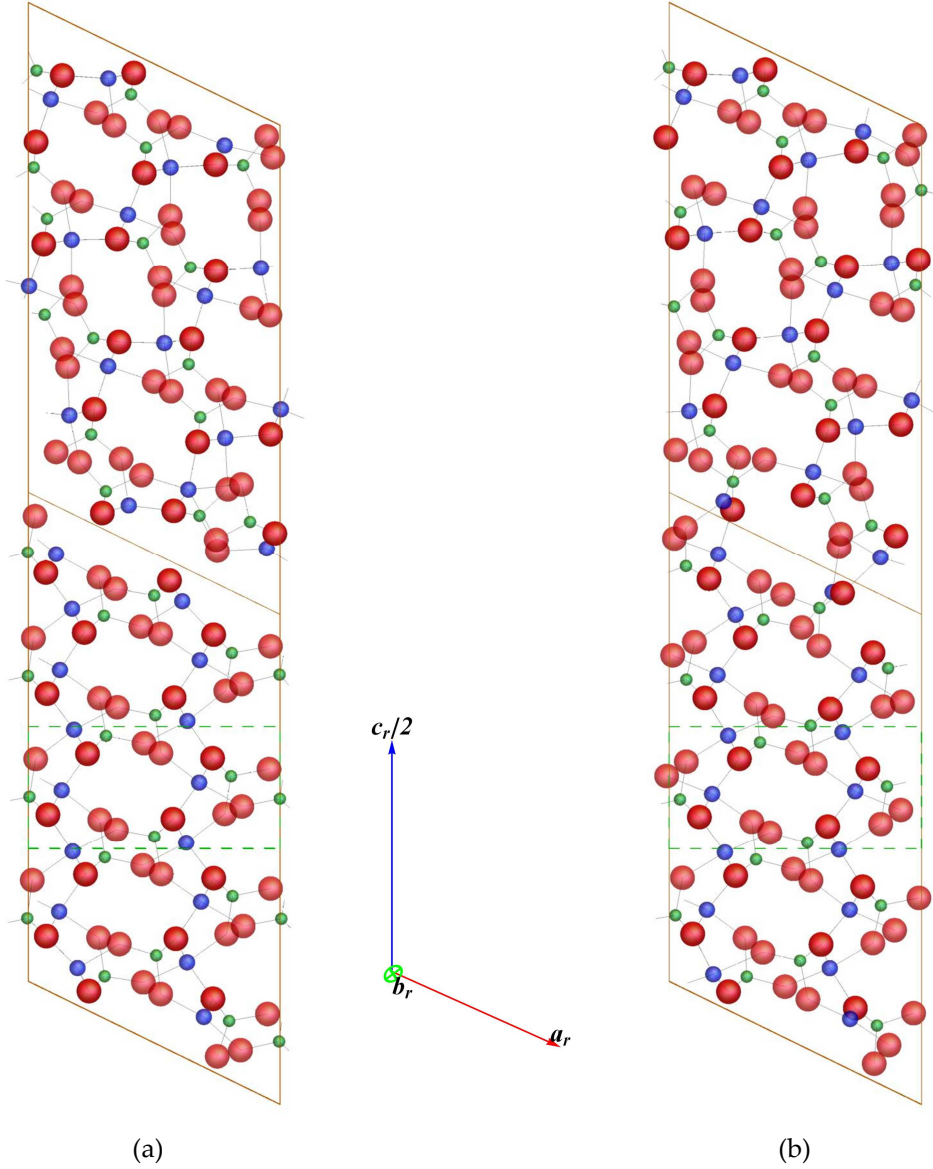

(a) (b)

**Figure 3.** $\Sigma$ 4 (101)/[010] GB for FP: (**a**) configuration 1 ; (**b**) FP configuration 2. The color scheme is the same as in Figure 1, including the FP unit cell and supercell translation vectors.

Let us now discuss other properties of the studied GB in the FP material even if they should be considered as preliminary. Bader charge analysis shows that for the bulk cell charges are as follows: $q(Fe) = +1.76e$, $q(P) = +3.55e$, $q(O(1)) = -1.32e$, $q(O(2)) = -1.34e$, and $q(O(3)) = -1.32e$. The sum of charges in the formula unit, *i.e.* $q(Fe) + q(P) + q(O(1)) + q(O(2)) + 2q(O(3))$, results in $+0.01e$, which is again practically zero taking into account rounding errors. With the charge neutrality being fulfilled, one can observe that the FP system becomes more covalent bonded, compared to LFP (see also discussion in Ref. [5]), since the magnitude of ionic charges diminishes (except for Fe assumed to be an $Fe^{3+}$ ion); the phosphate group has the charge $-1.76e$. This indicates that in LFP, the Li electron – supposed to be transferred solely to the Fe ion – is partially transferred also to the phosphate group (thereby oxygen anions). The more open FP structure with respect to LFP (see Figures 1 and 3) is also an indication of more covalent bonds in FP. In the case of the C1 GB, average ionic charges change slightly only ($0.01e – 0.02e$), but Fe ions close to the interface have a lower charge by about $0.04e$ (the largest changes occur for Fe ions with 4-fold coordination). The C2 GB behaves in about the same way as the C1 GB concerning ionic charges, but for Fe ions with 4-fold coordination the charge decreases even by $0.11e$ in average compared to the reference bulk system. Clusters of atoms were also investigated, but for FP GB configurations we chose somewhat different approach because of a less regular GB structure compared to LFP. Namely, 24 atom clusters (four formula units) were considered (instead of 6 atom clusters) forming just one unit cell of $FePO_4$. Such 24 atom clusters are 'parallel' with the GB interface and are located on either side of the interfaces and also in the bulk-like regions. There are totally 8 clusters in the supercell of each configuration. For both cases, there is a charge modulation going along the last supercell dimension. At the interfaces, one side is charged positively and the other one negatively. The magnitudes of cluster charges peak at $0.10e$ for C1 and $0.03e$ for C2. The overall charge is, of course, zero since the boxes have neutral charge. Again, the GB C2 interface with a better coherence exhibits lower cluster charges.

Magnetic moments of all atomic species in the reference bulk box are $\mu(Fe) = 4.01\mu_B$, $\mu(P) = 0.01\mu_B$, $\mu(O(1)) = 0.18\mu_B$, $\mu(O(2)) = 0.16\mu_B$, and $\mu(O(3)) = 0.00\mu_B$ (magnitudes are given). Because of an AF order, the total magnetic moment of the cell is zero. The magnetic moment of iron ions is slightly increased over the value for bulk LFP, but it does not reach the value $5\mu_B$ expected for a free $Fe^{3+}$ ion in the high spin state. As for GB configurations, magnetic moments of Fe atoms in configuration 1 are modified slightly only – usually within $0.02\mu_B$ – regardless of atomic positions (close to the interface or in bulk-like regions). The magnetic moments of oxygen atoms are also affected when they are close to the interface (and have dangling bonds). The size of the total magnetic moment is $0.08\mu_B$, indicating slight disturbances in the AF order. An examination of configuration 2 reveals less expected behavior. Namely, three Fe ions close to the interfaces have their magnetic moments diminished to about $3.33\mu_B$, while remaining Fe ions are almost unchanged. O ions with dangling bonds (*i.e.* those close to interfaces) have their magnetic moments also changed compared to bulk-like regions. The total magnetic moment of C2 is $1.16\mu_B$, suggesting a strong disturbance of the AF order. We mention that a GGA+*U* approach (see *e.g.* Ref. [33]) applied to the Fe $3d$ electrons can help to increase slightly Fe ionic charge and magnetic moment in the FP compound.

Concerning the positron behavior at the studied C1 and C2 configurations, preliminary calculations carried out using the so called atomic superposition method [34] – being the part of our computational approach [22] -- indicate that positrons are trapped at the grain boundary studied. The estimated lifetimes exceed the bulk lifetime by about 40 ps and 10 ps for C1 and C2 configurations, respectively. This is an interesting observation since there is no excess free volume for the examined boxes. Still the structure at the GBs provides more interstitial space compared to bulk (as seen in Figure 3), which explains this effect. The positron density (not shown here) shows only a maximum at the middle GB for both configurations, in contrast to LFP (see Figure 1). The calculated bulk FP lifetime amounts to 207.2 ps (GGA value) [28,29] and this value also reflects the more open structure of FP compared to LFP. Recent positron annihilation study of the FP system [35] suggests again that one lifetime component detected in measured spectra could correspond to positron annihilation at grain interfaces.

## 4. Conclusions and Outlook

The study of a special grain boundary in LFP ($LiFePO_4$) and FP ($FePO_4$) shows interesting and to some extent peculiar properties of GBs in these materials. In particular, we examined a near-CSL tilt $\Sigma 3$ (101)/[010] GB in LFP and its counterpart $\Sigma 4$ (101)/[010] GB in FP. For each GB type, two configurations (C1 and C2) were constructed with the aim to optimize the reconnection of atomic bonds at the GBs. In the case of $\Sigma 3$ GB (LFP), both configurations exhibit an acceptable coherence and C2 provides a lower GB energy. In addition, C1 GBs are slightly charged, whereas this effect is negligible for C2. Positron lifetime calculations suggest that both GB configurations trap positrons, which requires further experimental verification. GB excess free volumes do not correlate well with positron lifetimes, likely due to the complex morphology of the GB interstitial space. When Σ 4 GB (FP) is considered, the level of coherence at the GB interfaces decreases compared to LFP. We trace back this effect to changed (Σ 3 → Σ 4) GB coincidence conditions and modified character of the bonds in FP, which becomes a more covalent system when Li is removed from LFP. As a result, we observe a weak GB cohesion, slight charge modulation within the supercells and asymmetry in the GB structure between the two GB interfaces in the supercell. Nevertheless, the calculated ionic charges and magnetic moments are reasonable both in the bulk and in regions close to the interfaces. Positron lifetime calculations again suggest that positrons are trapped at the GB interfaces. Configuration C2 is regarded to have more regular properties than C1, including the fact that it possesses a lower GB energy. Thus, we expect that C2 has less influence than C1 on the performance and the degradation of LFP electrodes though it should be studied in more detail. On the other hand, both C1 and C2 trap positrons and most likely Li atoms as well. Therefore, this observation supports the hypothesis that both C1 and C2 lead to inhomogeneous Li distributions producing stresses and damages inside the cathode of LIBs. For the sake of completeness, it should also be mentioned that Lachal *et al.* [9] employed a chemical delithiation procedure, whereas in real LIBs electrochemical processes take place during delithiation, which might affect the GB behavior in somewhat different way.

In order to proceed with the Li diffusion study in the GB examined, we intend to improve the FP Σ 4 GB properties, *e.g.* by putting a few cations (Li or Fe) at the interfaces or shifting slightly the interface plane to become occupied by Fe cations. One could also attempt to relax a Σ 3 GB despite its poor coincidence properties. Independently, we plan to explore how the choice of exchange-correlation functional may influence the results of GB studies in LFP and FP materials. Specifically, GGA+*U* [33] and SCAN [36] functionals will be tested. Upon refining the characteristics of the examined FP GB, the affinity of Li to the GBs in LFP and FP will be checked to see the preferred Li ion position. As a next step, an *ab initio* molecular dynamics study will be undertaken. The Li diffusion mechanism is of primary interest – that is, whether Li ions can move just via one dimensional channels along the [010] direction or whether an interstitial mechanism is involved or other cation vacancies are needed for Li to move [37]. Such a study should help to answer the question how the GBs in LFP can affect the Li ion transport and especially if they can block it. In this way, DFT calculations and simulations provide a solid foundation to understand GB formation and the impact of this effect on the impedance and state of health of the battery [38]. Therefore, the current preparatory study motivates future research focusing on important GB issues affecting the battery performance and ageing.

**Author Contributions:** Conceptualization, J.K. and B.B.; Methodology, J.K.; Computations, A.P. and J.K.; validation, J.K.; writing—review and editing, J.K., A.P. and B.B.

**Funding:** J.K. was partially supported by the Czech Republic's Ministry of Education, Youth and Sports from the Large Infrastructures for Research, Experimental Development and Innovations project "IT4Innovations National Supercomputing Center – LM2015070". J.K. also appreciates the support by the Czech Science Foundation under project 17-17016S. B.B. acknowledges support from the COST Action CA16218.

**Acknowledgments:** The authors wish to acknowledge CSC – IT Center for Science, Finland for computational resources.

**Conflicts of Interest:** The authors declare no conflict of interest.